# STATISTICAL SENSOR FUSION OF A 9-DOF MEMS IMU FOR INDOOR NAVIGATION

J. C. K. Chow

Xsens Technologies, Pantheon 6a, 7521 PR Enschede, The Netherlands - jacky.chow@xsens.com

**Commission IV, WG IV/5**

**ABSTRACT:**

Sensor fusion of a MEMS IMU with a magnetometer is a popular system design, because such 9-DoF (degrees of freedom) systems are capable of achieving drift-free 3D orientation tracking. However, these systems are often vulnerable to ambient magnetic distortions and lack useful position information; in the absence of external position aiding (e.g. satellite/ultra-wideband positioning systems) the dead-reckoned position accuracy from a 9-DoF MEMS IMU deteriorates rapidly due to unmodelled errors. Positioning information is valuable in many satellite-denied geomatics applications (e.g. indoor navigation, location-based services, etc.). This paper proposes an improved 9-DoF IMU indoor pose tracking method using batch optimization. By adopting a robust in-situ user self-calibration approach to model the systematic errors of the accelerometer, gyroscope, and magnetometer simultaneously in a tightly-coupled post-processed least-squares framework, the accuracy of the estimated trajectory from a 9-DoF MEMS IMU can be improved. Through a combination of relative magnetic measurement updates and a robust weight function, the method is able to tolerate a high level of magnetic distortions. The proposed auto-calibration method was tested in-use under various heterogeneous magnetic field conditions to mimic a person walking with the sensor in their pocket, a person checking their phone, and a person walking with a smartwatch. In these experiments, the presented algorithm improved the in-situ dead-reckoning orientation accuracy by 79.8 – 89.5% and the dead-reckoned positioning accuracy by 72.9 – 92.8%, thus reducing the relative positioning error from metre-level to decimetre-level after ten seconds of integration, without making assumptions about the user's dynamics.

## 1. INTRODUCTION

In markerless close-range photogrammetry or laser scanning, the imaging instrument is often moved between tripod locations to achieve complete coverage of an object. A registration process is then performed to combine the data. Such a non-linear estimation process can benefit from good initial pose information (e.g. better than 1 metre for translation and 10 degrees for rotation (Bae, 2009)), which an Inertial Measurement Unit (IMU) can provide. IMUs are self-contained instruments capable of measuring accurate relative poses when the systematic errors are modelled well.

To process the IMU data, conventional Kalman filter textbooks usually introduce the IMU mechanization equations as part of the dynamics model for real-time navigation. Alternatively, in this paper, IMU data will be treated as sensor measurements in a batch least-squares framework to obtain a globally smoothed navigation solution while compensating for relevant systematic errors simultaneously.

## 2. PROPOSED METHOD

Sometimes it is impossible to perform a dedicated IMU self-calibration before capturing data in the field (e.g. due to cost/time restrictions). Other times, the user might be interested in estimating the dynamic portion of the systematic errors (e.g. gyro bias drift). The presented method can be applied in-situ to improve the quality of the navigation solution by formulating the IMU navigation problem as a calibration problem. Given uncontrolled motion and frequent magnetic disturbances, a set of calibration parameters (e.g. accelerometer and gyroscope biases, soft-iron effects, and hard-iron effects) can be estimated to improve the navigation solution, even though they may not be applicable out-of-sample. The proposed self-calibration method was performed offline using a robust batch optimization technique. The novelty of the presented 9-degrees-of-freedom (DoF) IMU calibration method is the combination of the following:

- Statistical sensor fusion of 9-DoF IMU data for more accurate navigation by batch post-processing.
- Tightly-coupled joint calibration rather than sequential hierarchical calibration of accelerometers, gyroscopes, and magnetometers.
- Number of optimization variables is time invariant and only contains the calibration parameters; this eliminates efforts in rotation parameterization, datum definition, and deriving initial approximation for the navigation states.
- Assumes only piecewise local homogeneous magnetic field rather than a single global homogeneous field, which is more suitable for indoor applications.
- Automatic outlier detection for abrupt and gradual magnetic disturbances during optimization; unsupervised detection of regions with constant local magnetic field.
- Opportunistic zero change in velocity, inclination, and loop-closure position updates when deemed suitable for calibrating the MEMS IMU indoors.

### 2.1 Sensor Models

The following sensor models describe the relationship between the observed signals (e.g. sensed angular velocity and acceleration) and the true signal by accounting for the systematic and random errors. In the following equations, uppercase letters represent 3 by 3 matrices and lowercase letters represent 3 by 1 vectors. Furthermore, $S$ is a diagonal matrix and $N$ is a lower-triangular non-orthogonality matrix, where both are composed of three independent unknown parameters (Zhang et al., 2010). Although the effect of scale-factor error and axes non-orthogonalities can be merged into a single matrix, they are treated separately for the accelerometers and gyroscopes to be consistent with some literature (e.g. Syed et al., 2007).

**2.1.1 Accelerometers:**

$$^s y_a = S_a N_a\,^s a + b_a + \varepsilon_a \quad (1)$$

where, $^s y_a$ is the measured acceleration in sensor frame
$S_a$ is the linear accelerometer gain
$N_a$ is the accelerometer axes non-orthogonality
$^s a$ is the true acceleration in sensor frame
$b_a$ is the accelerometer bias
$\varepsilon_a$ is the accelerometer noise

**2.1.2 Gyroscopes**

$$y_\omega = S_\omega N_\omega R_\omega\,^s \omega + b_\omega + G_\omega\,^s a + \varepsilon_\omega \quad (2)$$

where, $y_\omega$ is the measured angular rate in gyroscope frame
$S_\omega$ is the gyroscope gain
$N_\omega$ is the gyroscope axes non-orthogonality
$R_\omega$ is the inter-triad mis-alignments between the accelerometers and gyroscopes
$^s \omega$ is the true angular rate in sensor frame
$b_\omega$ is the gyroscope bias
$G_\omega$ is the g-sensitivity
$\varepsilon_\omega$ is the gyroscope noise

**2.1.3 Magnetometers:**

$$y_m = D_m\,^s m + o_m + \varepsilon_m^* \quad (3)$$

where, $y_m$ is the measured magnetic field in magnetometer frame
$D_m$ is the soft-iron effects
$^s m$ is the true magnetic field in sensor frame
$o_m$ is the hard-iron effects
$\varepsilon_m$ is the magnetometer noise

*Note that $D_m$ and $o_m$ also convey the effect of magnetometer biases, gains, axes non-orthogonality, and inter-triad mis-alignments.

**2.2 Constraints and/or Measurement Updates**

The constraints and measurement updates in this section are written in implicit form (i.e. f(X,Y) = 0, where X and Y are the unknowns and observations, respectively). This was chosen to eliminate the necessity to explicitly solve for the navigation states, which grows linearly with time.

**2.2.1 Accelerometers**: When static (or quasi-static) periods in the IMU data are detected, the magnitude of the measured acceleration in sensor frame (*s*) should equal the local gravity ($^L g$).

$$^s a_x^2 + {}^s a_y^2 + {}^s a_z^2 - {}^L g^2 = 0 \quad (4)$$

**2.2.2 Gyroscope**: When static (or quasi-static) periods in the IMU data are detected, the magnitude of the measured angular rate should equal the rotation rate of the Earth. However, for MEMS IMUs the noise typically masks such a weak signal. Instead, the three components (*x*, *y*, and *z* channel) can be conditioned to be zero.

$$^s \omega_x = {}^s \omega_y = {}^s \omega_z = 0 \quad (5)$$

**2.2.3 Gyroscopes + Magnetometers:** To separate the signal originating from the movement of the sensor from the ambient signals (e.g. gravity), the gyroscope signals and magnetometer signals can be compared to each other in both static and dynamic situations. Instead of requiring the sensor to be static, the local magnetic field should be constant and homogeneous. If this assumption is satisfied, the magnetometer can act as a low-pass filter that smooths out the sensed angular rate, while the gyroscope captures the high-frequency dynamics missed by the magnetometers. Although such an assumption may be satisfied in outdoor applications, in indoor urban environments the constant and homogeneous magnetic field assumption is often violated. Instead, the assumption can be relaxed to assume only piecewise constant and homogeneous magnetic fields at the expense of losing the absolute heading reference (i.e. magnetic north) and experiencing possible heading drift.

Assuming a longer duration of homogeneous magnetic field can reduce heading drift and induce more smoothing; however it is more likely to be violated due to magnetic disturbances. On the contrary, by assuming a shorter duration, the update has a higher probability of being valid but the heading will drift more rapidly. It has also been perceived that a longer duration assumption is more robust to magnetic disturbances because with a larger rotation interval the outliers become more detectable. To combine the benefits of both approaches, the magnetometer updates can be performed at two frequencies simultaneously (e.g. 100Hz and 10Hz). The magnetic field measurements at different times can be related through a 3D rotation (Equation 6) determined by performing strap-down integration on the gyroscope signal (Equation 7).

$$dq_{t,t+T} \cdot {}^s m_{t+T} \cdot dq^c_{t,t+T} - {}^s m_t = 0 \quad (6)$$

$$dq_{t,t+T} = \frac{1}{2} \int_{\tau=t}^{T} dq_{t,\tau} \cdot {}^s \omega_\tau d\tau \quad (7)$$

where, $dq_{t,t+T}$ is the relative change in orientation from time $t+T$ to time $t$ expressed using quaternions

**2.2.4 Accelerometers + Gyroscopes:** Several updates based on the accelerometer and gyroscope can be enforced depending on the user's motion for calibrating the sensors, namely levelling update, zero-velocity update (ZUPT), and coordinate update (CUPT).

**Levelling Update**
When the sensor is static the accelerometers can define the tilt angles relative to the local-level frame by measuring gravity. This can be used to update the inclination determined from integrating the gyroscope readings and give the orientations an absolute vertical reference.

$$dq_{t_1,t_i} \cdot a_{t_i} \cdot dq^c_{t_1,t_i} - a_{t_1} = 0 \quad (8)$$

**ZUPT**

Between two static (or quasi-static) periods the total change in velocity is zero. Instead of directly applying the update to the velocity, which would involve solving for all the nuisance intermediate velocity and orientation parameters, this update can be applied directly to the accelerometer and gyroscope signals using Equation 9.

$$^sg_{t_1} \cdot (t_i - t_1) + dv_{t_1,t_i} = 0$$
$$dv_{t_1,t_i} = \int_{\tau=t_1}^{t_i} dq_{t_1,\tau} \cdot {}^s a_\tau \cdot dq^c_{t_1,\tau} d\tau \qquad (9)$$

where, $dv_{t_1,t_i}$ is the change in velocity from time $t_1$ (first epoch) to time $t_i$ expressed in the sensor frame at $t_1$

**CUPT**

If the user is rotating the sensor while standing or sitting at the same location, and/or the sensor was returned approximately to the same position after a period of time, an approximate zero change in position update can be applied. In the former case, this update can be applied at regular intervals, and in the latter, it can be applied opportunistically even if the sensor is not static. Following the strap-down integration approach described above and assuming the sensor started at rest, the CUPT can be implemented without explicitly solving for the navigation states, as shown below.

$$\frac{1}{2} \cdot {}^s g_{t_1} \cdot (t_j - t_1)^2 + dp_{t_1,t_j} = 0$$
$$dp_{t_1,t_j} = \int_{\tau=t_1}^{t_j} \int_{\xi=0}^{\tau} dq_{t_1,\xi} \cdot {}^s a_\xi \cdot dq^c_{t_1,\xi} d\xi d\tau \qquad (10)$$

where, $dp_{t_1,t_j}$ is the change in position from time $t_1$ to time $t_j$ expressed in the sensor frame at $t_1$

### 2.3 Visual Representation of the Updates and Constraints

To summarize the different information used for calibration, a graphical representation of all the measurement updates working in coherence is shown in Figures 1 and 2 (as an example). Figure 1 provides a close-up view of the interaction of the observations with the constraints/updates and Figure 2 shows a more global view (of a different possible configuration). The orange circles are the calibrated sensor inputs, the purple circles are constant values (e.g. gravity when the sensor is static), the green circle is the strap-down integrated rotation quantity, the blue box is the approximately zero change in position update, and the red box is the zero change in velocity update. The constraints and updates may happen at different times (i.e. opportunistically), and only the magnetometer updates are being applied at regular intervals.

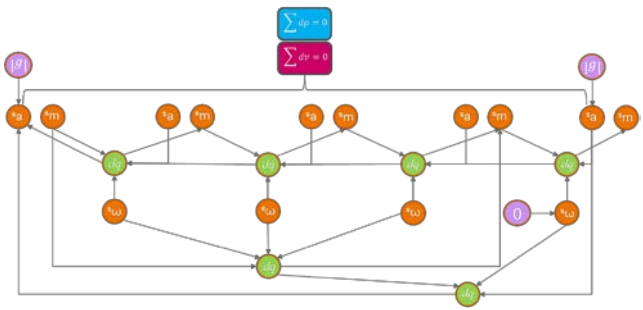

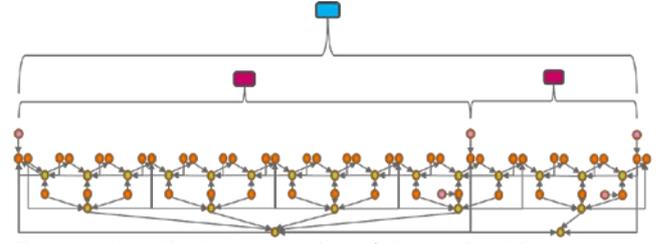

Figure 1: Graphical representation of the relationships between the observations (orange), known constant values (purple), and strap-down integrated quantities (red, green, and blue) on a local scale.

Figure 2: Graphical representation of the relationships between the observations, constraints, and updates on a global scale.

### 2.4 M-Estimator with L2 Regularization

The calibration parameters for the accelerometer, gyroscope, and magnetometers can be solved simultaneously in an iterative implicit least-squares adjustment (Förstner and Wrobel, 2004). The unknowns vector contains the 42 calibration parameters only (i.e. $X = [b_a, S_a, N_a, b_\omega, S_\omega, N_\omega, G_\omega, R_\omega, D_m, o_m]^T$) and the observations are the accelerometer, gyroscope, and magnetometer readings, along with the pseudo-measurements for the approximate CUPT. L2 regularization is applied to all the calibration parameters to prevent overfitting in cases where the parameters are unobservable under the in-situ dynamics. The nuisance parameters (i.e. states) were marginalized away by design from the beginning (i.e. in the functional models), which constrain the dimensions of the Hessian matrix. Through marginalization the Hessian loses its sparse structure; however, the size of the resulting dense matrix will never exceed 42 by 42 regardless of the amount of data captured. This is a favourable property and offers the potential for this method to be scaled to larger datasets.

The standard deviations for the sensors can be obtained from the manufacturer's specification sheets or from studying the Allan Variance. The noise introduced for the approximate CUPT depends on the application and requires tuning (it was set to 10 cm in this paper). All other measurement updates were assumed to be exact. To improve the robustness of the estimations, the Huber and Tukey weight functions (Zhang, 1997) were adopted for the accelerometer and magnetometer observations, respectively. This approach is suitable for detecting abrupt magnetic disturbances that significantly affect individual observations, for example. For more gradual changes in the magnetic field, a single residual may happen to fall below the detection threshold; in that case residuals of consecutive measurements can be tested together using the generalized likelihood ratio test (Gustafsson, 2010). This was applied to the magnetometer residuals using a sliding window approach post-adjustment. Afterwards, the batch optimization using the M-estimator is repeated with the outliers downweighted. The navigation solution can be determined efficiently using the IMU mechanization equations with the adjusted accelerometer and gyroscope observations after compensating for the systematic errors.

### 3. EXPERIMENTATION

Two MEMS-based IMUs from Xsens Technologies, MTi-300 and MTi-G-700, both with built-in accelerometers, gyroscopes, and magnetometers were used for testing the algorithm. All data was logged at 100Hz (note: the raw inertial data were captured at 2kHz and then down-sampled to 100Hz via strapdown integration (Vydhyanathan et al., 2015)).

**3.1 Walking with IMU Close to the Torso**

MTi-G-700 data was captured in a typical office environment with many magnetic objects. As shown in Figure 3, there was little rotation in the first scenario (with the changes in heading being the most pronounced) and the measured magnetic field is far from being homogeneous. The calibrated results with and without applying CUPT ($\sigma_{CUPT}$ = 10 cm) is given in Figures 4 and 5, respectively. It should be noted that because the ZUPT can be applied with better precision than the CUPT, the benefit of including loop-closure in the calibration is only minor in this trial. This also suggests that it is not crucial for the user to revisit previous locations during the in-situ calibration, which is advantageous. However, if the adjusted tracking solution is desired and the loop-closure can be detected reliably, applying it can improve the estimated trajectory (Figure 6). From Table 1 it can be observed that the errors with and without CUPT post-calibration are comparable. In both cases the accelerometer and gyroscope errors were reduced to about 0.09 m/s$^2$ and 0.16 deg/s, which is similar to the results from a dedicated calibration performed on-site that had more dynamic movements to generate samples with a better spread over the sensor's measurement ranges. With the inclusion of the approximate loop-closure update, the errors in the integrated acceleration were slightly higher, but this was more than compensated by the improvement in accuracy of the orientation estimate, which dominates the strap-down integration performance.

Although the in-sample error for the in-situ calibration is comparable to the on-site calibration, the in-situ calibration results are highly correlated with the motion; therefore, its calibration parameters are less transferable to out-of-sample trials than a calibration that has stronger excitations about every axis. For instance, the recovered hard-iron parameters from the on-site calibration was [-0.0191, 0.0152, -0.0130]$^T$, while the recovered hard-iron effect from the in-situ calibration was significantly different, [-0.7740, -0.8649, 0.0693]$^T$. Nonetheless, the calibration improved the overall accuracy of the strap-down integrated navigation parameters by approximately 90% for the in-sample motion.

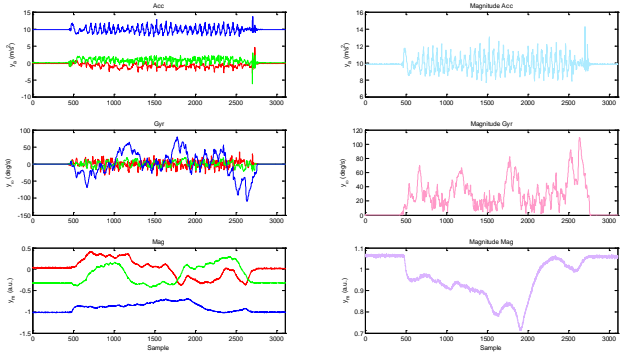

Figure 3: Input signal from the walking trial (with the IMU near the torso) that was used for in-use calibration. The red, green, and blue in the left column represents the x, y, and z components, respectively.

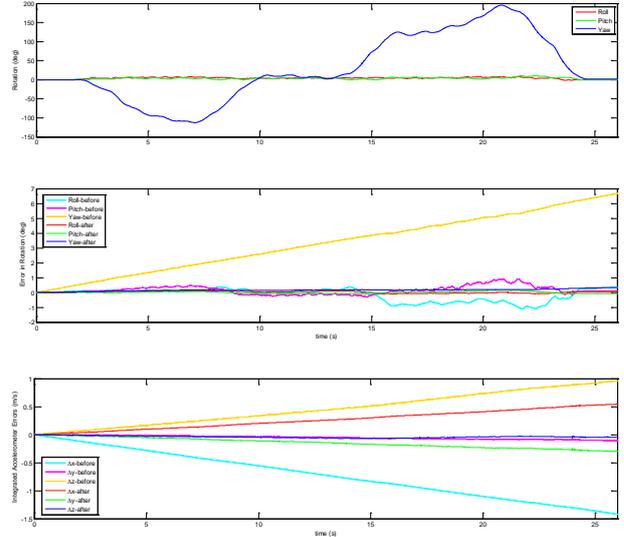

Figure 4: (Top) the reference orientation of the walking trial (with the IMU near the torso) determined by iMAR; (Middle) orientation errors from integrating the gyroscope signal; (Bottom) accelerometer errors from integrating the accelerometer signal in a rotating frame before and after in-situ self-calibration with CUPT.

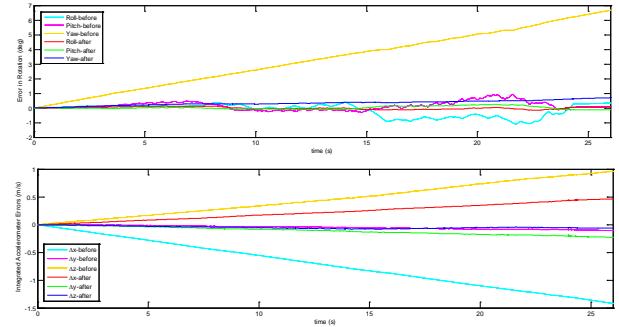

Figure 5: (Top) orientation errors from integrating the gyroscope signal; (Bottom) accelerometer errors from integrating the accelerometer signal in a rotating frame before and after in-situ self-calibration without CUPT.

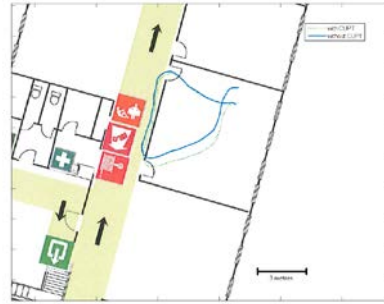

Figure 6: Estimated trajectory of the IMU post-calibration with and without applying loop-closure

|  | Before | After - with CUPT | | After - without CUPT | |
|---|---|---|---|---|---|
|  | Error | Error | % Improv. | Error | % Improv. |
| Acceleration (m/s$^2$) | 0.10 | 0.09 | 6.70 | 0.09 | 7.00 |
| Angular | 0.22 | 0.16 | 26.40 | 0.16 | 26.11 |
| Rate (deg/s) | | | | | |
| Orientation (deg) | 2.23 | 0.12 | 94.67 | 0.23 | 89.52 |
| Velocity @ 10s (m/s) | 0.85 | 0.05 | 94.70 | 0.07 | 91.60 |
| Position @ 10s (m) | 2.92 | 0.24 | 91.68 | 0.21 | 92.76 |
| Integrated Acceleration (m/s) | 0.99 | 0.36 | 64.02 | 0.30 | 69.85 |

Table 1: RMSE of the input signals and estimated navigation quantities before and after in-situ user self-calibration with and without apply loop-closure to the walking trial (with the IMU near the torso).

**3.2 Picking Up and Replacing the IMU – Checking Smartphone/Tablet**

The MTi-300 and iMAR were sitting still on an office desk until being picked up by the user. Following a few mocked keystrokes and finger swiping motions the IMUs were returned approximately to the same area on the desk. The calibrated IMU signals of the MTi-300 are shown in Figure 7. It can be seen from these figures that the amount of motion is limited and the sensor transitioned between two different magnetic fields. The first is caused by the table and the other is the field in front of a computer while the user is looking at the sensor. The orientation of the sensor during the trial along with the orientation errors and integrated acceleration errors when compared to the iMAR is provided in Figure 8. The in-sample error is reduced by applying the self-calibration method. Using the set of updated calibration parameters resulted in over 85% accuracy improvement in both the integrated angular rate and integrated acceleration (Table 2).

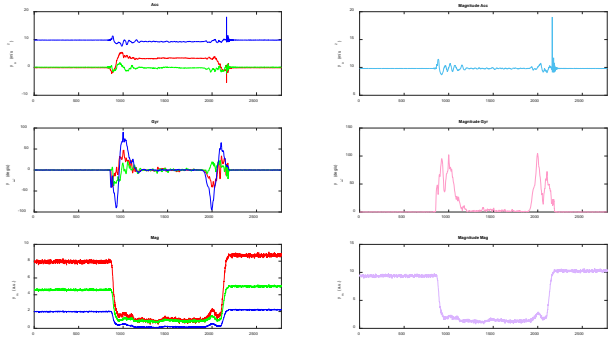

Figure 7: Input signals for in-use IMU self-calibration. The IMU was treated as a smartphone or tablet. The user reached for it on the table, and returning it after pretending to check for some messages. The red, green, and blue in the left column represents the x, y, and z components, respectively.

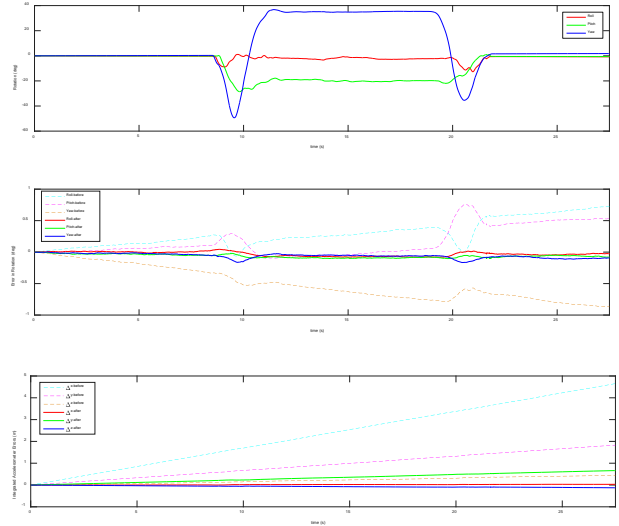

Figure 8: (Top) the reference orientation of the checking smartphone/tablet trial determined by iMAR; (Middle) orientation errors from integrating the gyroscope signal; (Bottom) accelerometer errors from integrating the accelerometer signal in a rotating frame before and after in-situ self-calibration with CUPT.

| | Before | After - with CUPT | |
|---|---|---|---|
| | Error | Error | % Improv. |
| Acceleration (m/s$^2$) | 0.12 | 0.06 | 52.48 |
| Angular Rate (deg/s) | 0.11 | 0.11 | 3.13 |
| Orientation (deg) | 0.43 | 0.06 | 85.94 |
| Velocity @ 10s (m/s) | 0.19 | 0.07 | 61.46 |
| Position @ 10s (m) | 0.83 | 0.22 | 72.91 |
| Integrated Acceleration (m/s) | 2.94 | 0.35 | 88.03 |

Table 2: RMSE of the input signals and estimated navigation quantities before and after in-situ user self-calibration for the picking up and replacing phone trial.

**3.3 Walking with IMU Close to the Wrist – Smartwatches**

The periodicity of the signal shown in Figure 9 is caused by the arm swinging motion while walking. Although the repetitive pattern of walking can be exploited to provide additional information to the MTi-300 calibration, only the measurement updates and constraints described in the Proposed Methods section of this paper were used. When walking around a typical office the ambient magnetic field is often changing due to objects such as fire extinguishers, electronic devices, shelves, and cabinets. Despite the challenging heterogeneous magnetic field conditions, the sensor's in-sample accuracy was improved by determining an updated set of calibration parameters using self-calibration as shown in Figure 10. From Table 3 it can be observed that the errors found in the integrated accelerometer and gyroscope signals were greatly reduced, yielding approximately 90% and 80% improvements, respectively.

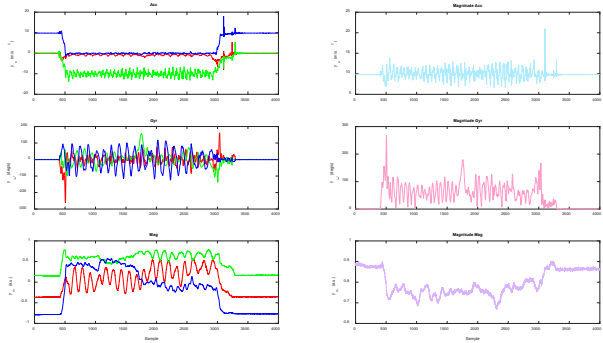

Figure 9: Input signal from the walking trial (with the IMU close to the wrist) that was used for in-use calibration. The red, green, and blue in the left column represents the x, y, and z components, respectively.

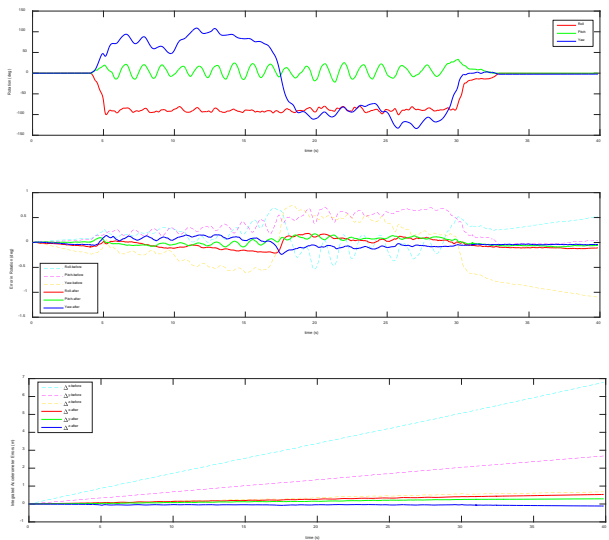

Figure 10: (Top) the reference orientation of the walking trial (with the IMU close to the wrist) determined by iMAR; (Middle) orientation errors from integrating the gyroscope signal; (Bottom) accelerometer errors from integrating the accelerometer signal in a rotating frame before and after in-situ self-calibration with CUPT.

|  | Before | After - with CUPT | |
|---|---|---|---|
|  | Error | Error | % Improv. |
| Acceleration ($m/s^2$) | 0.15 | 0.11 | 30.32 |
| Angular Rate (deg/s) | 0.13 | 0.13 | 3.46 |
| Orientation (deg) | 0.41 | 0.08 | 79.80 |
| Velocity @ 10s (m/s) | 1.24 | 0.13 | 89.21 |
| Position @ 10s (m) | 3.56 | 0.59 | 83.53 |
| Integrated Acceleration (m/s) | 4.42 | 0.36 | 91.79 |

Table 3: RMSE of the input signals and estimated navigation quantities before and after in-situ user self-calibration for the walking trial (with the IMU close to the wrist).

## 4. CONCLUSION

Calibration can add tremendous value to MEMS IMUs but it is an expensive quality assurance procedure that needs to be updated frequently. This paper presented a new total-system user self-calibration routine for a 9-DoF MEMS IMU. In contrast to calibrating the individual components separately, all sensors were jointly estimated to take advantage of their correlations. It encompasses static and dynamic inertial and magnetic information and applies attitude update, CUPT, and ZUPT without external equipment to estimate the calibration parameters. This self-calibration method was designed to be robust against inhomogeneity in the ambient magnetic field. Results have shown that the calibration quality does not deteriorate significantly in the presence of magnetic disturbances. The calibration was tested in-use under various heterogeneous magnetic field conditions with few excitations, mimicking a person walking with the sensor on the torso, a person checking their phone, and a person walking with a smartwatch. In the best case scenario, the presented algorithm improved the in-situ dead-reckoning orientation accuracy by approximately 90% and greatly reduced the positioning error (at approximately the mid-point between two positioning updates, the dead-reckoned positioning accuracy was improved by about 90%). Future work will focus on a near real-time implementation of this self-calibration method for in-use applications.

## ACKNOWLEDGEMENTS

This research is funded by TRAcking in complex sensor systems (TRAX), under the EU's Seventh Framework Programme (grant agreement No. 607400), and the Natural Science and Engineering Research Council (NSERC) of Canada. Valuable discussions with Jeroen Hol, Henk Luinge, Alexander Young, Giovanni Bellusci, Ignazio Aleo, Laurens Slot, and Matteo Giuberti are gratefully acknowledged.